\documentclass[12pt]{article}
\usepackage{graphicx} 
\usepackage{epstopdf}

\newcommand{\blankline}{\vskip .3cm}
\newcommand{\fa}{\begin{eqnarray}}
\newcommand{\ffa}{\end{eqnarray}}
\newcommand{\f}{\begin{equation}}
\newcommand{\ff}{\end{equation}}

\begin{document}
\centerline{\LARGE Conservative solutions to the black hole information problem}
\blankline
\blankline
\rm
\centerline{ Sabine Hossenfelder and Lee Smolin}
\blankline

\blankline
\centerline{\it Perimeter Institute for Theoretical Physics}
\centerline{\it Waterloo, Canada  N2J 2W9}
\centerline{\it  \ \ and }
\centerline{\it Department of Physics, University of Waterloo}

\blankline
\blankline
\blankline
\centerline{\today}
\blankline
\blankline
\blankline

\begin{abstract}
We review the different options for resolution of the black hole loss of information problem. We classify them first into
radical options, which require a quantum theory of gravity which has large deviations from semi-classical physics on
macroscopic scales, such as non-locality or endowing horizons with special properties not seen in the semi-classical
approximation, and conservative options, which do not need such help.  Among the conservative options, we argue that 
restoring unitary evolution relies on elimination of singularities.   We argue that this should hold also in
the AdS/CFT correspondence.

\end{abstract}

\tableofcontents

\newpage

\section{Introduction}

The conjecture that information is lost, and unitarity fails, when black holes evaporate, has challenged theorists since Hawking
first proposed it in 1974 \cite{Hawk74, Hawking:1976ra}.  After more than three decades of discussion, there have been many proposals, but no consensus on a solution.  In this 
comment, we would like to contribute to the solution of the problem by applying some basic logic to the different options on the table.  
We will argue that by making two simple classifications of the possible solutions one can argue, even in the absence of
a complete quantum theory of gravity, what the plausible resolution is.   

The first classification has to do with how heroic or radical are the steps asked of the quantum theory of gravity. We will label
a proposed solution to the information loss puzzle {\it radical} if 

\begin{itemize}

\item{} It attributes to the horizon or apparent horizon of a black hole physical properties which are not also properties of arbitrary 
null surfaces, or which are not apparent in a semiclassical treatment.  

\item{} It calls on extreme forms of non-locality in regions of spacetime with curvatures far below the Planck scale, or requires 
 transfer of quantum information over large spacelike intervals.  

\end{itemize}

If either or both of these things are true then the  quantum theory of
gravity attribute to weakly curved regions of space-time properties very different from those found in the semiclassical
description.  Any approach to resolution of the black hole information puzzle that does not make either of these assumptions will be called
{\it conservative.} 

There have been claims made that string theory has mechanisms that can underlie a radical solution of the problem \cite{BHC}, though the relevance
of these effects has been debated in \cite{ABHC}. We think it is fair to say that the questions of  whether or not string theory requires or admits
a radical solution to the information loss problem are presently open.

We would then like to advocate the view that before trying to make the radical solutions work we should make sure that an acceptable 
resolution does not come out of a conservative approach.   To begin with, it is not terribly plausible that the horizon would  
be the source of the 
resolution to the problem, because in classical and semi-classical General Relativity, the horizon is locally no different from any
other null surface.  It is only what happens globally, that determines if a surface is an horizon or not.  Because of this, it seems very unlikely that a quantum theory of gravity will do much to modify the physics near horizons.  At most one can expect a little bit of quantum
fluctuations present at the horizon to induce a {\it quantum ergosphere}, as discussed by York in \cite{york}.  But this is not enough,
by itself, to solve the black hole information puzzle.

But if we don't have reason to expect much from the horizon, the one place we do expect a quantum theory of gravity to have a major impact is on the singularity and the region approaching it, where the curvature invariants exceed Planck scales.  In fact, the idea that quantum gravity will remove the spacelike singularities of black holes and cosmological models is very old and supported by calculations in a variety of models and approaches \cite{ Frolov:1988vj,Frolov:1989pf,sing-go}.

In this paper we will not consider the question of whether particular approaches to quantum gravity resolve singularities.  We will, instead, assume that some theory does remove singularities and ask whether by doing so it leads to a resolution of the black hole information puzzle.  But we do note that there is recent work that does show that, in a particular model of quantum gravity, black hole singularities are removed in a way that leads to the restoration of unitary evolution.  This result has been derived in a study of the CGHS \cite{CGHS} model by Ashtekar et al in \cite{Ashtekar:2008jd}.  They find results that confirm earlier arguments in \cite{Abhay-Martin}.  
Part of the motivation of this paper is to put their results in a broader context. 

   Let us consider the evolution, within a quantum theory of gravity, of the spacetime of a star on the verge of
collapse to a black hole.  We can call the result a quantum black hole spacetime.  This kind of initial condition can be specified entirely within a semi-classical region.  We can be sure that such a spacetime will have a region of trapped surfaces, and hence an apparent horizon.   In such a context we can ask two questions, the answers to which lead to a classification of the conservative options for a resolution of Hawking's puzzle:

\begin{enumerate}

\item{} Is there a real event horizon?

\item{} Is there a singularity?

\end{enumerate}

Each of these has two possible answers, yes and no.  

We will first need to give more precise definitions of the terms involved. Once this is done we may consider the four possibilities that come from the possible answers to these two questions.  We will see that with minimal assumptions about the quantum theory of gravity, and more precise definitions of the terms involved, we can reach a simple and robust conclusion:

\begin{itemize}

\item{} Unitarity restoration in a conservative context requires a quantum theory of gravity that eliminates singularities. 

\end{itemize}

We will also argue below, that in a quantum theory of gravity where probability is well defined, and therefore conserved,
the absence of singularities plausibly implies there is unitary evolution. This then complements the results of Ashtekar et al \cite{Ashtekar:2008jd} which shows that in a specific model the singularity is eliminated and unitary evolution is restored.

In this paper we strive to make assumptions about the quantum theory of gravity only to the extent they are needed to frame
the initial paradox of Hawking and a resolution of it.  Thus, while we will give criteria for the presence of horizons and the absence of singularities, we make these the weakest criteria needed to draw definite conclusions.  As such, we expect  our conclusions to hold
in a large variety of theories and contexts. We do not aim to provide a concise review of the literature or previously
discussed solutions. Nor do we claim the point of view advocated here is original, our aim is to support an approach to resolving  the problem that seems to have been underappreciated\footnote{Other authors who proposed conservative solutions
include \cite{conservative,Hsu:2006pe}.}.  For reviews on the black hole information loss problem the interested reader is referred to \cite{Page,Preskill}.

In the next section we give some essential definitions, which makes it possible to classify the different options for a conservative
resolution of the black hole information problem in section 3.  In section 4 we discuss a few other issues, including objections to
remnants and baby universes and the extension of our arguments to the AdS/CFT context. $m_{\rm{Pl}}$ and $l_{{\rm Pl}}$ denote the
Planck mass and Planck length respectively.

\section{Definitions: asymptotic structures for quantum spacetimes}

We will start with defining some necessary terminology. Our aim is to make statements about a space-time with an initially 
semi-classical state whose evolution however might have a quantum gravitational phase or region. We will not define what a quantum 
spacetime is, as that depends on the approach to quantum gravity, about which we do not want to make any specific
assumptions. We are interested only
in a definition of quantum spacetimes which are semi-classical in the following sense:

\begin{quotation}
A quantum spacetime ${\cal Q}{\cal S}{\cal T}$ is partly semi-classical if 
\begin{itemize}
\item[a)] There is a procedure to define a manifold ${\cal R}$, a metric operator $\hat{g}_{ab}$ and a state $|\Psi \rangle $, such that $({\cal R}, \langle \Psi | \hat{g}_{ab} | \Psi \rangle )$ defines 
a possibly extendible Lorentzian spacetime (which we call for short $({\cal R}, \langle g_{ab} \rangle )$).
\item[b)] The semi-classical Einstein equations for $\langle g_{ab} \rangle$ are satisfied on ${\cal R}$, up to small quantum  corrections.  

\end{itemize}  \end{quotation}

Here, $({\cal R}, \langle g_{ab} \rangle ) $ may not be complete, ie there may be regions of ${\cal Q}{\cal S}{\cal T}$ that do not have a classical or semi-classical description.  If the quantum theory of gravity still allows us to define a manifold ${\cal M}$ such that
${\cal R} \subset {\cal M} $, then the ${\cal X}= {\cal M} -{\cal R} $ constitutes the {\it quantum region} of the quantum spacetime, while ${\cal R}$ constitutes its {\it semi-classical region} and   $({\cal R}, \langle g_{ab} \rangle ) $ is its {\it classical approximation.}

Since the  ${\cal Q}{\cal S}{\cal T}$ could have several semi-classical regions, we further need

\begin{quotation}

The classical approximation of a quantum spacetime is {\it semi-classically complete} if it contains all the regions of spacetimes that can be defined by the procedure that defines $({\cal R}, \langle {g}_{ab} \rangle )$.   

\end{quotation}

We can then make three important definitions

\begin{quotation}

A partly semi-classical quantum spacetime ${\cal Q}{\cal S}{\cal T}$ is {\it asymptotically flat} if its classical approximation
$({\cal R}, \langle g_{ab} \rangle ) $ is asymptotically flat.  

\end{quotation}

Depending on the quantum theory of gravity, a quantum spacetime ${\cal Q}{\cal S}{\cal T}$ may or may not have a causal structure that extends that of its  classical approximation. If it does not then the causal structure of a quantum spacetime is the causal structure of its classical approximation, otherwise the latter is a sub partial order of the former.  In either case, we can make several key definitions. 

\begin{quotation} 

A region of a quantum spacetime ${\cal Q}{\cal S}{\cal T}$ that is not in the causal past of ${\cal I}^+$ of its classical approximation $({\cal R}, \langle g_{ab} \rangle ) $ is an {\it asymptotically future hidden region}.

\end{quotation}

We can then say

\begin{quotation}
A quantum spacetime ${\cal Q}{\cal S}{\cal T}$ has a {\it future event horizon} if it contains an asymptotically future hidden region.
\end{quotation}

Note that we did not define the horizon itself, but merely a condition for its presence. We do this since we can not make
statements about the quantum gravitational regions. These definitions are well defined whether or not there is a quantum causal structure that extends the causal structure of its classical approximation.  The same is true of the following. 

\begin{quotation}
A {\it complete spacelike surface} $\Sigma$ of a quantum spacetime is a set of events of which no two are time- or lightlike to each other, 
and to which no events can be added without violating this condition.
\end{quotation}

If the causal structure does not extend past the classical approximation, then a spacelike surface is contained within the semi-classical region, otherwise it may extend.

Given such a complete spacelike surface we can linearize the quantum gravity theory (together with any matter fields) in its neighborhood and define a quantum theory of small fluctuations around $\Sigma$. We will assume that associated to any complete  spacelike surface $\Sigma$ of a quantum spacetime  there is an operator algebra
${\cal A}_\Sigma $ of local fields that describe small fluctuations in fields around the state $|\Psi \rangle $, which has a 
representation in a Hilbert space ${\cal H}_\Sigma $. 

We do not necessarily know what a singularity of a quantum spacetime is, but we can define a condition equivalent to the absence of any singularities:

\begin{quotation}

The quantum spacetime  is {\it quantum non-singular} if for any two complete non-intersecting spacelike hypersurfaces $\Sigma_1$ and $\Sigma_2$, there is a reversible linear map ${\cal M}_{1,2} : {\cal H}_1 \rightarrow {\cal H}_2 $, corresponding to dynamical evolution.   
Otherwise we say that the quantum spacetime is {\it quantum singular}.

\end{quotation}

Note that any slice means any slice in any slicing. The motivation for this definition is the following. 
The spacelike singularities in black holes that are the source of information loss in 
classical and semi-classical General Relativity do so by disabling the possibility for time reversal -- specifically time reversal in the theory of
small fluctuations propagating on a background spacetime.  If spacetime has a classical singularity it is geodesically incomplete \cite{HawkingEllis}, 
meaning there exists at least one time- or lightlike geodesics that can not be continued through the singularity. For this geodesic that ends at the singularity in a final point, 
consider a complete spacelike slice that meets the singularity in the same point. Whatever the information of the particle was that hit the singularity
on this incomplete geodesic, it does not get passed on to the spacelike slice\footnote{If spacetime is time-or lightlike geodesically incomplete it is in principle possible the spacelike slice can be continued through the singularity. This does not affect the argument since
it is only relevant the geodesic can not be continued.}. The above definition is a generalization of this classical definition without the
explicit referral to geodesics, since it is a notion that might not be well-defined in the quantum gravitational region.

One way to think about the loss of information in black hole evaporation is that less information is present after the evaporation in Hawking's original scenario because some of the information from ${\cal I}^-$ goes into the singularity and does not get out to ${\cal I}^+$. But looking at the time reversed situation this is not the case, all the information near ${\cal I}^+$ can be obtained from ${\cal I}^-$. Now
${\cal I}^{\pm}$ are not spacelike hypersurfaces but they represent the fact that in either slicing one will eventually 
(potentially at $t \to \pm \infty$) have to face the presence of the singularity. We could then propose the slightly 
different criteria for non-singularity that there is a reversible map between the Hilbert spaces for incoming and outgoing modes of a massless fields constructed on ${\cal I}^+$ and ${\cal I}^-$.  The definition we have given implies this in the
case of asymptotically flat spacetimes, because, in the absence of a classical singularity, ${\cal I}^+$ and ${\cal I}^-$ are each 
the limits of complete spacelike surfaces.  But the definition we have given is more general and, as we will see, applicable to the 
case of asymptotically AdS spacetimes.  

So we propose that a sufficient indication that no singularity is present is that the time evolution can be
reversed: if one can propagate information forward from $\Sigma_1$ to $\Sigma_2$, one can propagate it backwards from $\Sigma_2$ back to  $\Sigma_1$. 

Note that we do not require either that some curvature invariants blow up or some condition of geodesic incompleteness. What is needed is only that the propagation of the field modes between any two such surfaces can be used to define a reversible  transformation between 
the  Hilbert spaces associated with those surfaces.   This is good because we do not know if a particular quantum theory of gravity will allow us to define either curvature or geodesics in the quantum region. But we expect that if there is a singularity, in either of those senses, the condition we have just given of non-singularity can not be satisfied.  Since the latter can be defined more generally it is appropriate our purposes. We will discuss the relation to curvature singularities in section \ref{curvsing}.

Now reversibility is we may note, weaker than unitarity; it is a necessary, but not a sufficient requirement for unitarity. 
But we can argue that if the theory of small fluctuations propagating on the quantum spacetime has a sensible definition of quantum probability, then the map ${\cal M}_{1,2}$ will 
also be unitary.  An heuristic argument
for this goes as follows. If necessary, impose infrared and ultraviolet cutoffs so that the Hilbert spaces ${\cal H}_1$ and ${\cal H}_2$ are finite dimensional.  Since there is an invertible map between them they must have the same dimension, $n$.  Then 
the linear map ${\cal M}_{1,2}$ is an element of $GL(n,C)$ with non-vanishing determinant.  If we however impose the plausible requirement that
probability as determined by the inner products, is preserved in any basis, then ${\cal M}_{1,2}$ is an element of $U(n)$.  

\begin{figure}[th]
\centering \includegraphics[width=4cm]{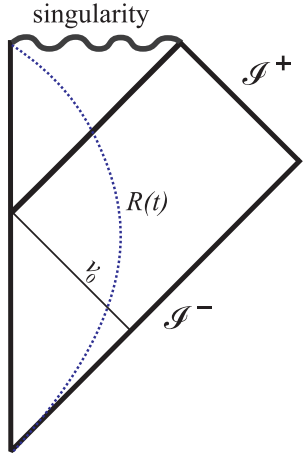}

\caption{Conformal Penrose diagram for the formation of a classical black hole.  The dotted (blue) line indicates
the surface of the collapsing matter with radius $R(t)$. The solid (black) line is the event horizon, the thin (black) line
labeled with $\nu_0$ is the last ray of light that escapes the fate of falling into
the singularity. \label{class}}
\end{figure}

This can be seen in the well studied example of particle production in time-dependent backgrounds, reviewed in \cite{BD,Ford:1997hb}.  
Consider that the fluctuations in question are described by a scalar field
$\hat{\phi} $ satisfying the usual massless scalar field equation, $\nabla_a \nabla^a \hat{\phi}=0$ in the semi-classical regions.  Assume also that  there is a time coordinate, $t_1$,  in the neighborhood of $\Sigma_1$, such that $\Sigma_1$ is defined by $t_1=$ constant,  and similarly 
for $\Sigma_2$.  Then we can follow the usual construction in which field
operators are decomposed into creation and annihilation operators based on positive and negative frequencies defined by $t_1$
and $t_2$, and construct the associated Fock spaces. We assume that the modes of $\hat{\phi} $ propagate between any two spacelike surfaces, whether 
semi-classical or not, so that,  ${\cal M}_{1,2}$ is a Bogoliubov transformation between
the creation and annihilation operators in ${\cal A}_{\Sigma_1}$ and ${\cal A}_{\Sigma_2}$. Given the completeness 
of the two surfaces, if there are no singularities where information of the modes might end up, then 
a unitary transformation
between ${\cal H}_{\Sigma_1}$ and ${\cal H}_{\Sigma_2}$ can be constructed from the Bogoliubov transformation \cite{DeWitt}. 
On the other hand, if  one
considers an incomplete slice as final region, the evolution will generically be non-unitary and evolve pure into mixed states.

The final defintion we need is that of  a quantum black hole spacetime.  The following will be 
sufficient for our purposes.

\begin{quotation}

A partially semi-classical, asymptotically flat quantum spacetime ${\cal Q}{\cal S}{\cal T}$ is a {\it quantum black hole  spacetime} if it has a complete spacelike slice in the semi-classical region with initial conditions that would 
classically lead to the formation of a horizon and a black hole.  

\end{quotation}

Assuming the standard positive energy conditions, were there no quantum effects the formation of  an horizon and a singularity, in the senses we have defined here, would be inevitable and result in the classical case of black hole formation as depicted in Figure \ref{class}. The question, to which we now turn, is what can happen in the quantum case.  

We will in the following restrict our attention to non-rotating, uncharged black holes.

\section{The options for the fate of quantum black holes}
\label{options}

For the discussion of this paper we will assume the existence of a quantum theory of gravity for which the definitions we have just given make sense. There are four logically possible scenarios to the evolution, which we now describe.

\subsection*{Option 1: Hawking's scenario: there are both horizons and singularities}

\addcontentsline{toc}{subsection}{Option 1: Hawking's scenario: there are both horizons and singularities}

Hawking's original argument \cite{Hawk74,Hawking} for a loss of information uses the following setting:

\begin{quotation}

{\bf Hawking's original scenario: } Let  ${\cal Q}{\cal S}{\cal T}$ be a quantum  black hole spacetime.  Then it will have an horizon and will be singular, according to the definitions above.  

\end{quotation}

\begin{figure}[ht]
\centering \includegraphics[width=5cm]{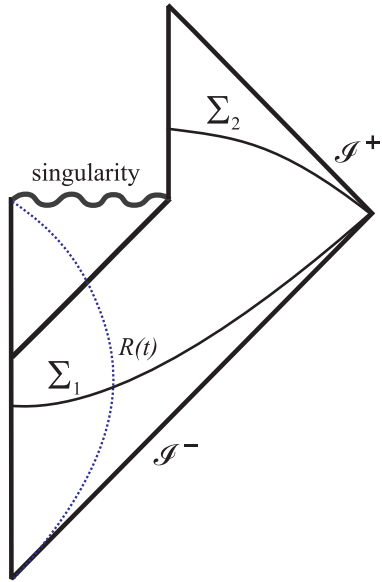}
\caption{Hawking's evaporation scenario.   The dotted (blue) line indicates
the surface of the collapsing matter with radius $R(t)$. The solid (black) line is the event horizon, the thin (black) lines labeled with $\Sigma_{1/2}$ 
are two complete spacelike hypersurfaces.\label{fig1}}
\end{figure}

We can refer to Figure \ref{fig1} for a view of what he had in mind.  In this case we say that unitary evolution has broken down and there is loss of information because there are two surfaces $\Sigma_1$ and $\Sigma_2$ such that there is no reversible map from ${\cal H}_1$ to ${\cal H}_2$.  No matter what the initial state was, the outcome is always a thermal state and it is not possible to find out from the final state what the initial state was.

Note that in Hawking's original argument the quantum theory of gravity retained the singularity of the 
classical and semi-classical black hole. As we will see shortly, one obtains a potential 
paradox only when this assumption is made.   We can conclude that Hawking was correct, in arguing that {\it if the black hole
spacetime retains an event horizon and spacelike singularity in the full quantum theory}, then information is lost and unitarity is
impossible. 

But let us see what the other options are. 

\subsection*{Option 2: Quantum naked singularities}
\addcontentsline{toc}{subsection}{Option 2: Quantum naked singularities}
\begin{quotation}

{\bf Quantum naked singularity scenario:} Let  ${\cal Q}{\cal S}{\cal T}$ be a quantum  black hole spacetime.  Then in the quantum theory it will be singular but have no event horizon.  

\end{quotation}

This is the worst possible outcome and, to our knowledge, no one is proposing it.  We mention it for completeness.  It would mean that whatever happens in the quantum region is not unitarily determinable by the past but is not hidden behind an horizon and  can affect the whole spacetime out to ${\cal I}^+$.  

Let us now discuss two ways that the information paradox could be resolved by a quantum theory.

\subsection*{Option 3: Neither horizon nor singularity}
\addcontentsline{toc}{subsection}{Option 3: Neither horizon nor singularity}

\begin{quotation}

{\bf Complete evaporation scenario:} Let  ${\cal Q}{\cal S}{\cal T}$ be a quantum  black hole spacetime.  Then in the quantum theory of gravity it will nonetheless be non-singular and it will be without an event horizon.  

\end{quotation}

\begin{figure}[ht]
\centering 
\includegraphics[width=5cm]{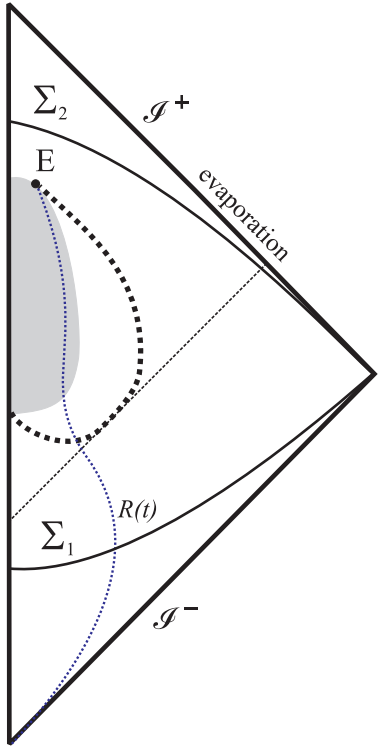}

\caption{Option 3: Complete evaporation scenario. The grey shaded region is potentially subject to non-negligible quantum gravitational corrections.
The thick dotted line is the apparent horizon, the thin dotted (blue) line indicates a possible world line of the surface of the collapsing
matter configuration. The thin dashed line represents the lightlike surface where the event horizon of the collapsing matter 
had been without evaporation. \label{fig2}}
\end{figure}

This is illustrated by Figure \ref{fig2}. In spite of there being an apparent horizon, indicated by the thick dotted line, there is not a 
real event horizon.  The apparent horizon ends at
point E, after which we can say that the black hole has evaporated. Plausibly, this would occur when the black
hole has shrunk down to Planckian size. Since the trapped surface could vanish in this quantum gravitational phase, it
might not be useful to think of this object as a black hole anymore. In this scenario, all the 
information that was trapped within the apparent horizon could eventually 
get out to infinity. Since the quantum spacetime is non-singular it follows that 
any complete spacelike surface $\Sigma_2$ to the future of E
has a state which is the image of a map from the pure state in ${\cal H}_1$ on the initial value surface $\Sigma_1$. The 
initial state thus can be recovered from the final one and there is no obstacle to there being a unitary map from a state on an initial surface to a state on a final surface.

This kind of scenario was discussed in \cite{Hasslacher:1980hd,Frolov:1981mz}, was later advocated by Ashtekar and Bojowald \cite{Abhay-Martin} and a detailed example of this kind of resolution was then provided by the treatment of the CGHS model \cite{CGHS} by Ashtekar et al in \cite{Ashtekar:2008jd}.
This is a $1+1$ dimensional model of quantum gravity coupled to a scalar field, motivated by string theory \cite{CGHSrev1,CGHSrev2}.  The classical
model has black hole solutions, and semi-classical calculations show that there is Hawking radiation and hence an information loss
puzzle. 
Ashtekar et al show in two controlled approximations that the singularity and horizon are eliminated in the quantum theory, leading to unitary evolution and a resolution of the information loss puzzle.

One key aspect of the Ashtekar et al work is that their methods allow them to study quantum spacetimes that have both regions which 
are semi-classical, AND regions which are far from classical,  characterized by very large fluctuations in expectation values of operators that measure the metric geometry.  The existence of quantum spacetimes which are in this way, non-uniformally approximated by semi-classical or effective field theory methods is
crucial to a resolution of the puzzle of black hole evaporation.  This is because without regions which are semi-classical, which include
asymptotic regions with timelike killing fields, the problems of unitarity and information loss can not be posed precisely enough to 
investigate.  That is, unless regions far from the black hole can be treated semi-classically, the context in which Hawking's puzzle arises can not be reproduced.  At the same time, the elimination of the singularity requires a region of space-time where the semi-classical
approximation fails badly. 
Arguments against this scenario have been offered on the basis of a claim that it would lead to long lived remnants. 
We will discuss these in section \ref{issues}. 

\subsection*{Option 4: There is an horizon but no singularity}
\addcontentsline{toc}{subsection}{Option 4: There is an horizon but no singularity}

A second way that quantum gravity could resolve the conjecture without loss of information is in the following scenario. 

\begin{quotation}

{\bf Massive remnant or baby universe scenarios:} Let  ${\cal Q}{\cal S}{\cal T}$ be a quantum  black hole spacetime.  Then in the quantum theory of gravity it will be non-singular, but there will still be an horizon.  

\end{quotation}

There are then two possibilities of such remnants:  

\begin{quotation}

{\bf 4A:} To the future of the event horizon, all complete spacelike hypersurfaces decompose into two regions that are not connected by any spacelike curves. There is then a region that pinches off from the asymptotically flat region, forming a new region of spacetime that we
will refer to as a {\it baby universe}. 

\end{quotation}

\begin{figure}[ht]
\centering \includegraphics[width=6cm]{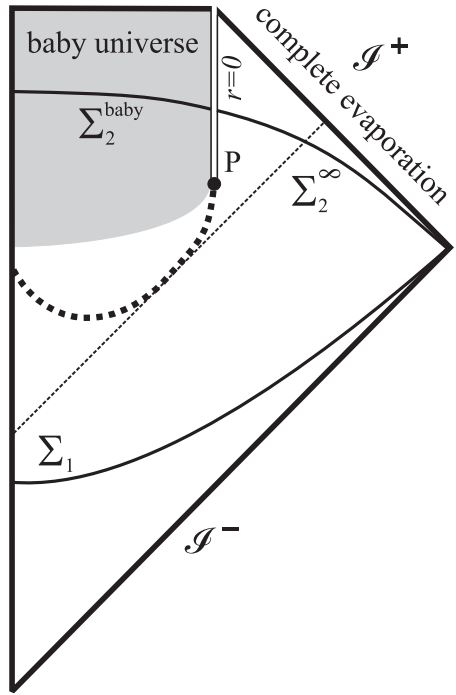}

\caption{Option 4A: Example of a  baby universe scenario. The pinch-off point is marked with P, the thick dotted line
indicates the apparent horizon. The double thin line is a boundary between two disconnected regions. The grey shaded 
region is potentially subject to non-negligible quantum gravitational corrections.  The thin dashed line represents the lightlike surface where the event horizon of the collapsing matter 
had been without evaporation. 
\label{fig3}}
\end{figure}

An example for this case is illustrated in Figure \ref{fig3}. The pinch off point is event P. There the 2-surfaces
making up the apparent horizon shrink down to zero area, but the volume contained within remains finite. The resulting  pinch off is illustrated in Figure \ref{fig3}. In this case there is no quantum singularity because there are complete surfaces, $\Sigma_2= \Sigma_2^{\infty} \cup \Sigma_2^{{\rm baby}} $  to the future of P which have two components, a component with an asymptotic boundary at spatial infinity and a compact component within the baby universe. 
There is no loss of information and no obstacle to unitary evolution, because the quantum information that falls behind the horizon survives indefinitely in the baby universe.  So the whole universe at a later time has a pure state, even if the component connected to 
${\cal I}^+$ can only be described by a density matrix.  This kind of resolution of the problem has been proposed by 
\cite{Frolov:1988vj,Frolov:1989pf,Hsu:2007dr,Hsu:2006pe}.

\begin{quotation}

{\bf 4B:} The horizon never shrinks down to zero size but stabilizes at a finite radius, forming a permanent massive remnant, surrounded by an event horizon. In contrast to scenario 4A, the spacelike surfaces do not fall apart into two
disconnected regions, and the remnant never disappears from the original spacetime. 

\end{quotation}

\begin{figure}[ht]
\centering \includegraphics[width=5cm]{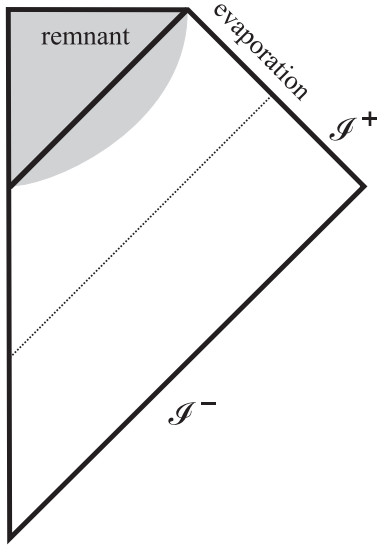}

\caption{Option 4B: Example for a massive remnant scenario. The solid line depicts the event horizon. The thin dashed line represents the lightlike surface where the event horizon of the collapsing matter 
had been without evaporation. The grey shaded 
region is potentially subject to non-negligible quantum gravitational corrections.  
\label{fig4}}
\end{figure}

In this case, unitary evolution can be maintained when the degrees of freedom in the interior of the remnant are taken into account. This is not that different in consequences from the baby universe scenario, as the spatial volume of the interior of the remnant may grow arbitrarily large (in comparison to $A^{\frac{3}{2}}$, where $A$ is the minimal area of the throat connecting the original spacetime to the remnant\footnote{Spherical symmetry allows an unambiguous definition
of the three-volume \cite{Wald}.}) while the exterior remains an horizon of fixed size. In contrast to option 4A, the massive remnant's ADM mass and its surface never reach
zero but they stabilize at a finite value. In either case unitary evolution is preserved, but there is a real horizon because the information trapped in the baby universe or remnant can never reach ${\cal I}^+$. This case is depicted in Figure \ref{fig4}. In both cases, 4A and 4B, the evolution
is unitary on complete hypersurfaces but will seem non-unitary in the exterior due to the incompleteness of the final region

There are some objections to these scenarios which we will discuss below.

\begin{figure}[ht]
\centering \includegraphics[width=11cm]{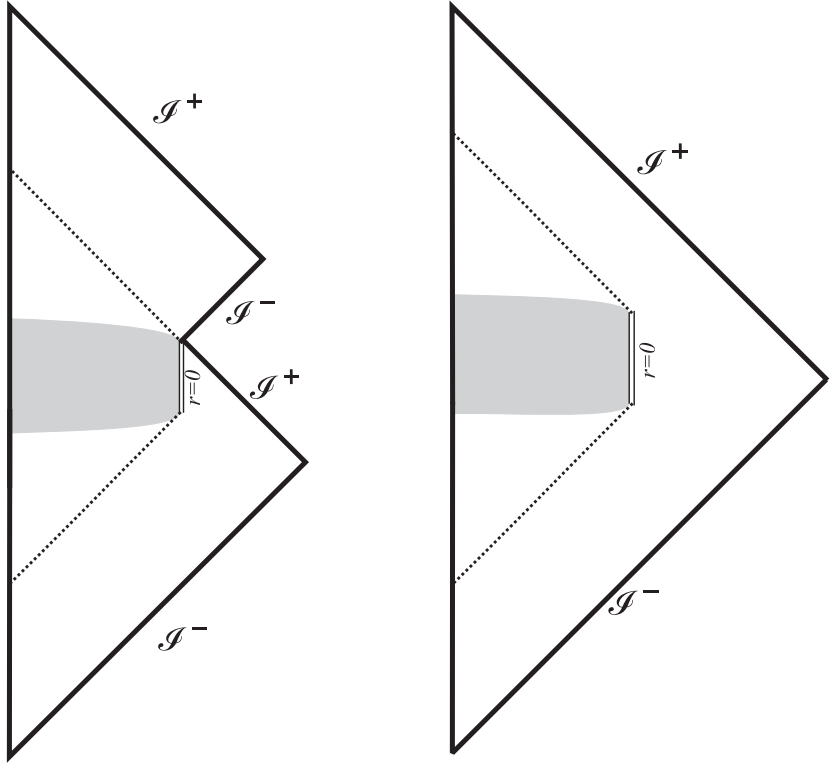}

\caption{Left: Option 4A: Example for baby universe scenario with a second asymptotically flat region. Right: Option 3: Example for an intermediate disconnected region that later reconnects. The double thin lines indicates a boundary between two disconnected regions. Grey shaded regions are potentially subject to non-negligible quantum gravitational corrections.
\label{2ar}}
\end{figure}

For completeness we note two variants of this option. To the future of the disconnected interior, there might follow another spacelike non-compact region that can be
asymptotically flat, an example is shown in Figure \ref{2ar}, left. Or, if the disconnected region reconnects to the
mother universe, we have an alteration of option 3, depicted in Figure \ref{2ar}, right. We emphasize however that though possible, both of these scenarios are
unlikely in a conservative approach, because they are not evolutions of complete initial data on ${\cal I}^-$.  In the first case ${\cal I}^-$ is not the complete initial data surface, in the second there is no dynamics known that could mediate the reconnection of disjoint regions
of quantum spacetime. 

In all of the above cases, we can not tell for certain how far the quantum gravitational regions will extend and
whether an intermediate quantum region might be followed by another semi-classical region. The grey shaded regions in all of these
figures indicate the possible presence of quantum gravitational effects, though they might indeed
be negligible in some of these areas. For example, if the interior of the baby universe grows very large, one would
expect that quantum gravitational effects only play a role at the junction of the mother- and baby universe to resolve the singularity.

\subsection*{The cost of saving quantum mechanics}
\addcontentsline{toc}{subsection}{The cost of saving quantum mechanics}

These four cases cover the possible options for a conservative resolution of the black hole information problem, ie those  in which the quantum black hole spacetime has or has not an horizon or singularity.
The first and second inevitably suffer from a loss of information, since the evolution is not time reversible, which means in particular the evolution of quantum states cannot be unitary. The third and fourth have reversible evolution. Reversibility is necessary, though not sufficient, for unitary evolution, but as we have argued earlier, a theory that has reversible but non-unitary evolution would lack a sensible definition of quantum probability.  

The  third and fourth scenario eliminate the classical singularity and so both have quantum information existing to the future of where the classical singularity would have been.  They differ only on whether an observer near ${\cal I}^+$ could recover that information.  If she can then there was not a real horizon, only an apparent one.  If she can't then the information still exists, but is somewhere she can never see it, either in a permanent remnant (4B) or in a new region of spacetime that has pinched off from hers and from now on evolves on its own (4A).  

Thus, the cost of saving quantum mechanics, without reliance on radical modifications of spacetime in the semi-classical domain, 
is either a baby universes, a remnant, or a quantum region 
that eventually leaks out its quantum information to infinity.  

If one wants or hopes that quantum theory is not be modified fundamentally by its incorporation of gravity, so that the resolution of the black hole information problem is conservative, then one wants one of these scenarios to work out. We see that the important thing to understand, to resolve the problem,  is what happens in the quantum gravitational phase and how the singularity is avoided in the quantum theory of gravity.

\section{Other issues}
\label{issues}

\subsection{Objections against stable, or quasi-stable remnants}
\label{remnants}

There is a literature critical of the option where there remain remnants, either permanent or long lived \cite{Giddings:1993vj,Giddings:1993km,Susskind:1995da}. If these are correct there remain two conservative options:  
a baby universe or a quick decay of the quantum region without a long lived remnant.  Nonetheless,  the 
arguments against remnants are not, in our view, definitive, for reasons
we would like to explain here. 

\subsubsection*{There is not enough information remaining inside a remnant}

There is a weak and a strong interpretation of the Bekenstein-Hawking entropy \cite{weak, weakandstrong,Sorkin:1997ja,Jacobson:1999mi}.  The weak form says that the black hole entropy $S_{{\rm BH}}= M^2$ (in Planck units) is a measure of the number of bits of quantum information that could be gotten by measurements made at the horizon.  Another way to say this is that the Bekenstein
bound is a measure of channel capacity for an apparent horizon.  
The strong form 
asserts that $S_{{\rm BH}}$ is the amount of quantum information needed to specify the state to the interior of the horizon, ie log of the dimension of the subspace of Hilbert space needed to describe the interior.  There are several reasons for ruling out the strong form, which are reviewed
in \cite{weakandstrong}.  

For example, if  one takes the strong form one then must consider the following problem.  A black hole formed in the past at mass $M_{\rm i}$ and then evaporated down to the present when its mass is $M_{\rm f}$. The semi-classical calculations asserts that the black hole has produced thermal radiation 
that does not carry any information (except its temperature). Thus, the 
information -- if preserved -- must remain within the black hole.  This requires a  larger set of possible interior states than the strong 
interpretation of the Bekenstein-Hawking entropy accounts for, as that  shrinks during the evaporation process to $S = M_{\rm f}^2$ while
the information content increases because it contains, in addition to the original state of the star that collapsed accross the horizon, the entangled partners of the Hawking radiation.  This is the inevitable consequence from a conservative
approach, and it applies while the black hole is still in the semi-classical regime\footnote{A similar argument is the `Hawking radiation cycle' discussed in \cite{Marolf:2008tx}.}.  The conclusion must be that the strong form of the Bekenstein-Hawking entropy is false and a weaker form is correct.  The only candidate we are aware of for a weaker form of black hole entropy is that given above. 

We further note that the strong form of the Bekenstein bound is explicitly violated by highly entropic states, considered in \cite{Marolf:2003wu} and  constructed in \cite{Hsu:2007dr}. There is in General Relativity no obstacle in
principle to solutions with arbitrarily small ADM mass and arbitrarily large entropy, and we have seen that they are
a necessary consequence of Hawking evaporation without modifications in the semi-classical regime. The presence of such objects does not
require any radical assumptions, whereas their avoidance does. It does thus seems puzzling why this option has
not received much more attention as a possible solution to the black hole information loss problem. One objection against
their existence is that they lead to singularities, but if singularities are removed by quantum dynamics then this cannot be an objection.
Another objection is that 
they are in conflict with the strong form of the Bekenstein entropy bound. But this conflict can be resolved by accepting the weak form of the Bekenstein-Hawking entropy\footnote{Another way to say this is that if one believes in the strong form of the Bekenstein-Hawking entropy and so has a problem with remnants, one has actually has a more serious problem.}. 

We next turn to the other often raised objection against stable or quasi-stable remnants.

\subsubsection*{Remnants must have an infinite cross section for pair production}

This argument begins with the assertion that effective field theory applies to remnants, so that they should, for purposes of physics above an length  scale $\lambda_c \gg l_{{\rm Pl}}$, be treated as point particles which can be created and annihilated and described by local field operators \cite{Giddings:1993vj,Giddings:1993km}. This is asserted to be the case because their physical diameter is much smaller
than $\lambda_c$.  However they must come in an infinite numbers of varieties because they have to contain the quantum information needed to restore the pure state of a thermal radiation produced by a potentially arbitrarily large initial black hole mass, as discussed above.  Since the couplings in effective field theory can not resolve any structure with a scale much below 
$\lambda_c$, each of these species must have the same pair production cross-section. Even if the pair production cross section for one of these species is exponentially suppressed, it is multiplied by an infinite factor, so that if they existed there would be an infinite pair production rate for remnants in arbitrarily weak background fields, as long as the available
energy exceeds two times the remnant mass.
\begin{figure}[ht]
\centering \includegraphics[width=11cm]{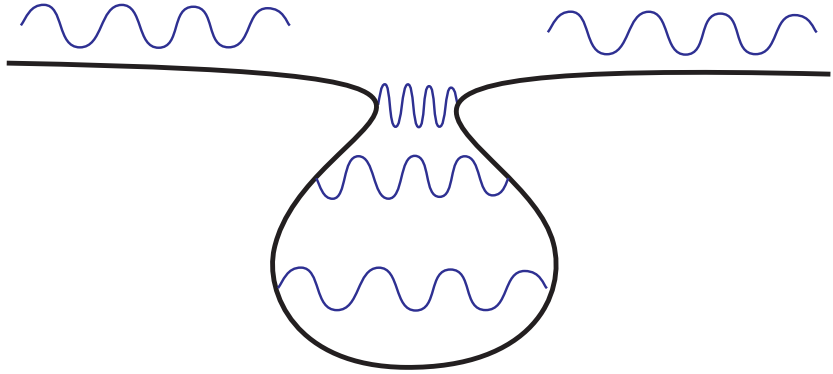}

\caption{Schematic picture of a ``bag of gold'', ie a remnant with a large internal volume, embedded in an asymptotically flat space.  
Shown is a mode $| k \rangle $ which is long wavelength
both far from 
and within the bag of gold to illustrate why it is implausible in an effective field theory long
wavelength modes outside the remnant decouple from degrees of freedom within the remnant. 
\label{bog}}
\end{figure}

The problem with this argument is twofold. 

First, it must be possible to define an effective field theory in this context.  This depends on making a separation between modes with
wavelength shorter and longer than $\lambda_c$, and integrating over the former to find an effective theory for the latter.  However note that a characteristic property of a black hole is the blueshift of frequencies when one approaches the horizon. This represents a
position dependent change for wavelengths that is specific to strongly curved backgrounds and is not a feature
of flat space quantum field theory. There is then no distinction between short and long wavelength modes that is valid in the whole spacetime.  

Related to this is our second point. Remnants are end stages of black hole evaporation, they are neither point particles nor are they in the semi-classical regime. They may be
very small on the outside, because they can be entirely contained within a small surface area. But their interior can be a throat that opens up into an arbitrarily large internal volume. As has been discussed in the literature, the interior of the remnant 
is a region that can have a volume large compared to $M_{\rm f}^3$ and indeed be growing. This is the ``bag of gold" scenario, discussed in \cite{bog}, and illustrated in Figure \ref{bog}. In this case, the excited modes of fields corresponding to whatever formed the black hole, whatever fell into it, and the unfortunate partner of each quanta of Hawking radiation, all end up here and as the interior expands will have very subplanckian wavelengths. 

An effective field theory approach to remnants would only be possible to the extent that 
 long wavelength modes with 
$\lambda > \lambda_c \gg l_{{\rm Pl}}$ decouple from the degrees of freedom in the interior of the remnant, which can distinguish the different possible internal states of the remnant.  However, it is doubtful this  is the case. Consider a mode such as that shown in Figure \ref{bog} which is very sub-Planckian in a region of space-time far from, or prior to,  the formation of the black hole, very high frequency near the apparent horizon, but again very long wavelength in the interior of the horizon.  Note that as it approaches the apparent horizon the mode $|k \rangle$ is very blue shifted, and it continues to blue shift as it comes closer to the region where the classical singularity would have been.  After it enters the expanding bag of gold region it can red shift again. If the bag of gold is expanding the mode in the interior can be redshifted again to arbitrarily long wavelength.

It is thus a long wavelength mode in the exterior region that can interact strongly with the degrees of freedom inside the remnant.  Its existence precludes the application of effective field theory to remnants.  Moreover, the existence of such modes makes it impossible to assume that the amplitude for interaction between a long wavelength external mode and a remnant is independent of the degrees of freedom in the interior of the remnant.  But this is a key assumption in the argument we sketched above for an infinite pair production rate for 
remnants\footnote{Related to this, one can argue that 
the interaction between any long wavelength mode and a highly entropic internal state should be suppressed the higher the entropy of the internal state.}. One might argue that in the presence of a horizon the interior is disconnected from the exterior. However, in a quantum theory, these regions are not disconnected as long as Hawking radiation, which can be understood as a tunneling process, is possible. A stable remnant does by assumption
no longer emit Hawking radiation, but whenever something falls in its mass increases and it will emit until it has reached the remnant mass again. 

There is in fact no more reason to think that a black hole remnant should be created by a local operator, than any other region of spacetime with an arbitrary large volume and high entropy. 
The geometry of the bag of gold to the interior of a remnant throat could be as large, weakly curved and have as many degrees of freedom and at least as much entropy as a star. Indeed, we would expect many black hole remnants to be the future evolutions of stars.

The situation is unchanged if the black hole is produced by quantum tunneling, rather than by classical gravitational collapse.  So long as the remnant does not completely pinch off, there is an event horizon that an external quanta's wavefunction may fall into.  
Thus, the case for the essential assumption needed to set up a description of a remnant by a single local operator in effective field theory, which is the 
decoupling of its internal degrees of freedom from long wave length degrees of freedom of the exterior, is not conclusive.
 If an effective description of such remnants 
is possible at all, their couplings should depend on their internal state even in the low energy limit. 
This possibility has been discussed for some
while \cite{Banks:1992mi,Banks:1992is,Banks:1994ph,Giddings:1993km,Giddings:1993vj,Giddings:1994qt} in the mid 90s but
no consensus appears to have been attained about it\footnote{We note that the inconclusiveness of these investigations has 
been pointed out already eg in \cite{Page, Jacobson:1999mi}.}.  

Other objections against baby universes have been addressed in \cite{Hsu:2006pe}.

\subsubsection*{Life time of decaying remnants}

If there is not a permanent remnant, then there arises the question how long the last stages of black hole evaporation will last. 
The Planck scale remains of the black hole may slowly decay, potentially exceeding the lifetime of the universe, or it may quickly dissolve on
microscopic timescales. We will here not advocate
a specific answer, but just mention several points of view.

A standard objection to the fast decay is that there is a large amount of quantum information, proportional to the entropy of the 
Hawking radiation, $\sim M_{{\rm i}}^2/m_{{\rm Pl}}^2$, where $M_{\rm i}$ is the initial mass of the collapsing matter, in an object of 
eventually very ADM small mass $\sim m_{{\rm Pl}}$.  At one quanta per bit, each quanta
will have a tiny amount of energy, $\sim m_{{\rm Pl}}^3 /M_{{\rm i}}^2$ and the more information, the less energy per bit one has. 
The emission of such a  small amount of energy takes a long time,  and since the bits of information come out of a very small region of 
spacetime they can not be emitted together because their wave-functions would overlap and spoil the encoding of information.

The only actual calculations for the resulting lifetime that we are aware of are \cite{remnantevap1} and \cite{remnantevap2}. Preskill 
\cite{Preskill} offers an heuristic argument based on the former references, and Giddings \cite{Giddings:1992hh} presents a similar estimate. Reference 
\cite{remnantevap1} however make use of the small volume
of the remnant and \cite{remnantevap2} uses a moving mirror analogy that presupposes the final
quantum gravitational object has still an horizon, neither of which might be applicable to the actual endstate.  

If one considers the recovery of information takes place only in the Planckian phase, as the most conservative
approach would indicate, this requires that the strong form of black hole entropy be false, as we discussed above. If one insists instead on the strong form then 
the black hole {\it must} leak information correlated with the thermal Hawking radiation, as the black hole evaporates, so that at any time, $t$, during the evaporation,  no more than $\sim \exp( M (t)^2) $ qbits are needed to restore the external state to a pure state.  On this assumption there is very little information in the black hole by the time it evaporates down to the Planck mass and the remains can quickly dissolve. However, the
emission of information in the Hawking radiation typically involves non-local physics in areas where quantum gravitational effects
are expected to be negligible to get the information from the quantum gravitational region to and/or through the horizon.

The scenario with leakage prior to the Planckian phase does seem to be indicated by the calculations of Ashtekar et al \cite{Ashtekar:2008jd}. It is also supported
by considerations of quantum information theory as argued by Hayden and Preskill \cite{Hayden:2007cs} -- although we note their argument 
assumes the strong form of black hole entropy meaning it assumes information can not stay inside and has to come out.  

Taken together we see again that it is our lacking understanding of the strong quantum gravitational effects near the
singularity that prohibit us to formulate a satisfactory solution to the information loss problem, and more investigation
is required to understand the properties of Planck-size, Planck-mass objects. 

\subsection{AdS/CFT}

Let us discuss the implications of the AdS/CFT conjecture from the point of view above.  To do this we have to state a 
couple of standard modifications of our earlier definitions. 

\begin{quotation}

A partly semi-classical quantum spacetime ${\cal Q}{\cal S}{\cal T}$ is {\it asymptotically AdS} if its classical approximation
$({\cal R}, \langle g_{ab} \rangle ) $ is asymptotically AdS.  

\end{quotation}

We note that the boundary of an asymptotically AdS spacetime is a timelike surface $\cal S$, of one dimension lower than the AdS space, on which there is
a flat Minkowski metric.  One can then define an asymptotic time coordinate, $t$, both in the asymptotically AdS metric and the Minkowski spacetime.  

\begin{quotation}

An horizon of an asymptotically AdS quantum spacetime ${\cal Q}{\cal S}{\cal T}$ is the boundary of  the causal past of $\cal S$ of its classical approximation $({\cal R}, \langle g_{ab} \rangle ) $.  

\end{quotation}

The AdS/CFT conjecture is then the following \cite{ads}.  

\begin{quotation}

{\bf AdS/CFT conjecture:}  There is a quantum theory of gravity ${\cal Q}{\cal G}$ which encompasses partly semi-classical asymptotically AdS spacetimes.  It has a Hilbert space ${\cal H}_{{\rm bulk}}$ and a Hamiltonian $H_{{\rm bulk}}$ which generates time evolution in the asymptotic
time coordinate $t$.  There is also a conformal field theory ${\cal Q}{\cal C}$ on $\cal S$ with a Hilbert space
${\cal H}_{{\rm boundary}}$ and Hamiltonian $H_{{\rm boundary}}$ that generates time evolution in the same $t$.  There is also another Hamiltonian in ${\cal H}_{{\rm boundary}}$, given by $H_{{\rm CFT}} \propto H_{{\rm boundary}} + {\cal D}$ where ${\cal D}$ is another generator of the conformal group acting on $\cal S$.  

These two theories are isomorphic to one another in the sense that there exists an  isomorphism ${\cal I}_{{\rm AdS/CFT}}$
\f
{\cal I}_{{\rm AdS/CFT}}(t) : {\cal H}_{{\rm bulk}} \leftrightarrow  {\cal H}_{{\rm boundary}} ,  
\ff
that takes the operator algebras to each other.
\f
{\cal I}_{{\rm AdS/CFT}}(t) :   {\cal A}_{{\rm bulk}} \leftrightarrow {\cal A}_{{\rm CFT}}
\ff
and in particular
\f
{\cal I}_{{\rm AdS/CFT}}(t) :   H_{{\rm bulk}} \leftrightarrow H_{{\rm CFT}}
\ff
such that time evolution commutes with ${\cal I}_{{\rm AdS/CFT}}(t)$ in the sense that
\f
{\cal I}_{{\rm AdS/CFT}}(t) \cdot e^{-\imath H_{{\rm CFT}} t } = e^{-\imath H_{{\rm bulk}} t }  \cdot {\cal I}_{{\rm AdS/CFT}}(t) . 
\ff

\end{quotation}

It follows from the statement of this conjecture that the ${\cal Q}{\cal G}$ should be quantum non-singular.  Given a complete spacelike surface $\Sigma$ in a semiclassial region of the bulk,  there should be an invertible 
inclusion map, ${\bf I}_\Sigma $, of the operator algebras, or density matrices such that pure states of the full theory restrict to pure states of the linearized theory.  

\fa
{\bf I}_\Sigma &:& {\cal A}_{\rm bulk} \rightarrow {\cal A}_{\Sigma}\\
{\bf I}_\Sigma &:& {\cal H}_{\rm bulk }  \rightarrow { \cal H}_{\Sigma}
\ffa

Now consider two such complete spacelike slices, $\Sigma_1$ and $\Sigma_2$, which intersect $\cal S$ at boundary times, 
$t_1$ and $t_2$.  Given a state $| \Psi_{\rm boundary}, t_1 \rangle$ in ${\cal H}_{\rm boundary}$,  we can map it with ${\cal I}_{{\rm AdS/CFT}}^{-1}$
to get a state $| \Psi_{\rm bulk}, t_1 \rangle$ in ${\cal H}_{{\rm bulk}}$.  
We can then use  $ {\bf I}_{\Sigma_1}$  to map this to a state in the Hilbert space of the linearized theory on $\Sigma_1 $, 
which is ${ \cal H}_{\Sigma_1}$.

We can also use $e^{-\imath H_{{\rm CFT}} (t_2-t_1)} $ to evolve $| \Psi_{\rm boundary}, t_1 \rangle$ to a state
$| \Psi_{\rm boundary}, t_2  \rangle$ in ${\cal H}_{{\rm boundary}}$ and then again use  ${\cal I}_{{\rm AdS/CFT}}^{-1}$ to map it to a pure
state $|\Psi_{\rm bulk}, t_2 \rangle$ in ${\cal H}_{{\rm bulk}}$.  We then use ${\bf I}_{\Sigma_2}$ to map this to a state in the Hilbert space of
the linearized theory  on $\Sigma_2 $, 
which is ${ \cal H}_{\Sigma_2}$.

Since $e^{-\imath H_{{\rm CFT}} (t_2-t_1) }$ is unitary and ${\cal I}_{{\rm AdS/CFT}}^{-1}$
is an isomorphism this gives a unitary map on from  $ { \cal H}_{\Sigma_1}$ to ${ \cal H}_{\Sigma_2}$. 
Thus, if the AdS/CFT conjecture is true, 
the evolution in the bulk between any two such hypersurfaces is unitary and there can not be information loss. There can 
however also not be a quantum singularity between these slices.  

Thus, if the assumptions of this argument are correct then we are left with scenarios three and four.

\subsection{Curvature singularities}
\label{curvsing}

Given a theory of quantum gravity, one could consider alternative notions of singular spacetimes. A more natural definition
than the one we have used here might be thought to be one in which 
a singularity occurs when expectation values of operators representing scalar 
invariants of the curvature tensor diverge. A quantum spacetime could then be said to be {\it curvature non-singular} 
if there is no event where the expectation value of a scalar function of the curvature tensor diverges. One then
could ask whether both notions of singular space-times coincide. Given that we do not know
the fundamental theory we can only make some general remarks about this. 

For spacetime to be quantum singular, there can be no two initial states that in the quantum spacetime
evolve into one and the same, and such arrive as the same state on the final hypersurface. For this to happen, it 
is however sufficient if there is an attractor for states that erases their differences. Without further 
knowledge about the relation between the spacetime and the quantum fields propagating in it, it is impossible 
to say whether this attractor must be accompanied by a divergence.

Similarly, whether or not a curvature singularity implies spacetime is quantum singular for the quantum field 
depends on the evolution laws coupling the spacetime to the propagating field. One could then have three 
different cases in which a curvature singularity would not imply a quantum singularity for the propagating quantum field:

\begin{enumerate}
\item Even though space-time has a curvature singularity, the quantum field has not, e.g. because the quantum
field does not resolve the structure, or gravity decouples in the high density limit.
\item Even though space-time has a curvature singularity no information runs into it, e.g. because it 
is redistributed\footnote{We assume that the no-cloning theorem holds and exclude that information can be copied.} 
and survives elsewhere. Since this means the information has to get out of the horizon before it
is gone, this typically implies non-local effects. 

\item The singularity itself can carry information and in such a way pass it on through the quantum region, this is 
realized
e.g. in the proposal by Horowitz and Maldacena \cite{Horowitz:2003he}. 
\end{enumerate}

\section{Discussion}
\label{discussion}

We would now like to make some comments on the debate on the information loss problem, in light of the conclusions we have reached.

What seems to us puzzling about much of the discussion in the literature is that the obvious conservative solution, that 
the singularity is removed and unitarity so restored, has not been given more attention. The information loss problem seems
paradoxical only if one believes it can not simply be blamed on our lacking understanding of the processes in the
quantum gravitational region. This is the case only if one accepts as true the arguments that have
been raised against the conservative solutions we have classified here, arguments which we do not believe are
definitive, precisely because we lack understanding about the quantum gravitational phase. If one believes these arguments 
however, then one has reason to construct more fanciful and radical solutions. 

Our main conclusion is that there is only need to invent radical solutions in the case that  the right 
quantum theory of gravity will not eliminate the 
singularity in Figures \ref{class} and \ref{fig1}.  
Were this the  case, then we would  apparently have to solve  problem of how to evolve unitarily between 
 ${\cal I}^-$ and ${\cal I}^+$ in Figures  \ref{fig1}.  This may lead us to  radical mechanisms 
such as black hole complementarity that imply non-local transfers of large amounts of quantum information 
over large space-like intervals \cite{BHC}, or to argue that a non-perturbative completion of gravity 
displays non-locality on the scale of the horizon \cite{Giddings:2007ie}. But that 
problem is moot if the actual causal structure of a quantum black hole spacetime looks like  Figure \ref{fig2} or \ref{fig3}. 

We then find it encouraging that one of the most  detailed quantum gravity calculation of the fate of an evaporating black hole that so far has been done shows that the singularity is restored and unitarity evolution between ${\cal I}^-$ and ${\cal I}^+$ is confirmed \cite{Ashtekar:2008jd}. This work shows that, at least in the example studied, once the singularity is eliminated, the causal structure of a black hole that forms from gravitational collapse and then evaporates is that of Minkowski spacetime, so that restoration of unitary evolution is almost trivial. 

Assuming the singularity is eliminated, there remains the question of whether 
 the information contained in the initial state returns to infinity, as in option 3 or remains trapped in a baby universe or permanent
 remnant as in option 4.  In the case of option 3 (which seems to be suggested by the results in \cite{Ashtekar:2008jd}), it is not
 enough to invoke unitarity, 
 one has still to understand how the information comes out.  So, not surprisingly,  more work is needed concerning the Planck scale regime, when the semiclassical approximation breaks down, but this is 
no reason to give up and retreat to radicalism. If one takes the conservative 
point of view that we have presented here, then an important issue is to  understand the 
properties of the highly entropic endstate of the black hole evaporation. We thus hope to encourage more
investigation of this subject.

\section{Conclusions}
\label{conc}

We have argued here that elimination of the singularity is sufficient to resolve the quantum information problem raised by Hawking in the context of black hole evaporation.  We reached this conclusion by noting that, in the absence of knowing the theory of quantum gravity, an appropriate condition for a quantum spacetime to be non-singular is the existence of reversible maps between Hilbert spaces describing the linearized quantum fluctuations of quantum
fields between spacelike surfaces. This then provides a useful definition of non-singularity to 
classify the options for resolving the information paradox without invoking departures from the semi-classical description outside
the region where quantum gravitational corrections can be expected to be strong. 

As we have seen, in the cases where the singularity is eliminated, no information loss can occur. The scenarios which resolve it are either that the black hole evaporates completely, or leaves a permanent massive remnant, or forms a baby universe. In the first case there is no real horizon, but in the latter cases there is.  Which is the actual resolution depends on the details of the quantum theory of gravity. Moreover, there is no reason in principle why a single theory could not have states where each of these options are realized, following the generic expectation that in a
quantum theory everything that can happen will happen unless explicitly forbidden. We further have seen that the arguments so far offered 
against short- and long-lived remnants or baby universes are not convincing due to a lacking investigation of the properties of highly
entropic objects with curvature in the Planckian regime.  

Finally, we considered the issue in the context of the AdS/CFT correspondence and reached the conclusion that the conjecture of 
equivalence of the bulk and boundary theory requires that the bulk quantum spacetime be non-singular, in the sense defined here. 

\section*{Acknowledgements}

We are grateful to Abhay Ashtekar, Steve Giddings, Stephen Hsu, Ted Jacobson and Don Page for conversations and correspondence on this issue.  
Research at Perimeter Institute for Theoretical Physics is supported in
part by the Government of Canada through NSERC and by the Province of
Ontario through MRI.

\end{document}